\documentclass[aps,prd,preprintnumbers,showpacs,nofootinbib,onecolumn,showpacs]{revtex4}

\usepackage{latexsym}
\usepackage{epsfig}
\usepackage{amssymb}
\usepackage{amsfonts}
\usepackage{amsmath}
\usepackage{amsthm}
\usepackage{marginnote}
\usepackage{color}
\newcommand{\lp}{\left(}
\newcommand{\rp}{\right)}
\newcommand{\lb}{\left[}
\newcommand{\rb}{\right]}
\newcommand{\al}{\alpha}
\newcommand{\bt}{\beta}
\newcommand{\ga}{\gamma}

\newcommand{\la}{\lambda}

\newcommand{\hGamma}{\hat{\Gamma}}
\newcommand{\hR}{\hat{R}}
\newcommand{\R}{{\mathcal R}}
\newcommand{\Lag}{{\mathcal L}}
\newcommand{\bg}{\bar{g}}
\newcommand{\bh}{\bar{h}}

\newcommand{\bGam}{\bar{\Gamma}}
\newcommand{\bnabla}{\bar{\nabla}}

\newcommand{\be}{\begin{equation}}
\newcommand{\ee}{\end{equation}}
\newcommand{\lsim}   {\mathrel{\mathop{\kern 0pt \rlap
  {\raise.2ex\hbox{$<$}}}
  \lower.9ex\hbox{\kern-.190em $\sim$}}}
\newcommand{\gsim}   {\mathrel{\mathop{\kern 0pt \rlap
  {\raise.2ex\hbox{$>$}}}
  \lower.9ex\hbox{\kern-.190em $\sim$}}}
\newcommand{\jose}[1]{#1}
\newcommand{\mk}[1]{#1}

\begin{document}

\title{The bimetric variational principle for General Relativity}
\author{Jose Beltr\'an Jim\'enez$^{1}$, Alexey Golovnev$^{2}$,
 Mindaugas Kar\v{c}iauskas$^{3}$ and
Tomi S. Koivisto$^{4}$}

\email{Jose.Beltran@unige.ch;  agolovnev@yandex.ru; mindaugas@ugr.es; t.s.koivisto@astro.uio.no}
\affiliation{$^1$D\'epartement de Physique Th\'eorique and Center for Astroparticle Physics,
Universit\'e de Gen\`eve, 24 quai Ansermet, CH--1211 Gen\`eve 4,
Switzerland\\
$^2$High energy physics department, St. Petersburg State University; Ulyanovskaya ul., d.1;\\ 198504 St. Petersburg, Stary Petergof; Russia \\
$^3$ CAFPE and Departamento de F\'isica Te\'orica y del Cosmos, Universidad
de Granada, Granada-18071, Spain \\
$^4$Institute of Theoretical Astrophysics, University of Oslo, 0315 Oslo, Norway \\}

\date{\today}

\begin{abstract}

The bimetric variational principle is a subtle reinterpretation of general relativity that assumes the spacetime connection
to be generated by an independent metric. Unlike the so called Palatini formalism that promotes the connection into a fundamental field, 
the new variational principle results in a physically distinct theory since the potential for the connection carries new degrees of freedom.  
The connection-generating metric naturally allows also an antisymmetric component. This sets torsion propagating! It \mk{is also shown here} that
while in the most straightforward generalization of the Einstein-Hilbert action the nonmetric degrees of freedom become ghosts, there exist very simple actions which give rise to viable theories \jose{at the linearised level} when subjected to the bimetric variational principle. \jose{However, the non linear interactions might bring unpleasant features like the Boulware-Deser ghost. This remains to be explored since this new type of bimetric theories does not, in principle, lie in the class of usual bimetric theories where non-linear interactions inevitably come in with new ghost-like degrees of freedom.}

\end{abstract}

\maketitle

\section{Introduction}

Gravity is unique among the fundamental interactions of Nature because, while electroweak and strong interactions occur between particles in a \jose{background}
spacetime, gravity describes the dynamics of the spacetime itself. The relation of gravity with the geometrical properties of spacetime lies at the heart of one of the most successful theories of physics, General \mk{R}elativity (GR). An issue of paramount importance is then to identify the degrees of
freedom associated with the spacetime geometry.

In GR, the dynamics of spacetime is described solely by its metric. However, from differential geometry we know that spacetime, when described as a manifold, can be endowed also with an affine structure. To formulate physical theories of gravitation, it is indeed necessary to introduce both metric and affine structures \cite{schrodinger1950space}. Firstly, the need for measuring distances between two spacetime points\footnote{Actually, the existence of a metric in a space is a stronger condition than the existence of a distance function, but we shall assume that such distance function is the one compatible with a scalar product.} or, equivalently, for measuring angles between vectors, forces us to introduce a metric structure. The affine connection is then needed to parallel transport vectors from one spacetime point to another. The concept of parallelicity is in principle independent of the metric structure. In GR however, the following relation is postulated:
\be \label{l-c}
\Gamma^{\al}_{\phantom{\al}\bt\ga}=\frac{1}{2}g^{\al\la}\lp g_{\bt\la,\ga}+ g_{\la\ga,\bt}-g_{\bt\ga,\la}\rp\,.
\ee
Thus the connection is set to be the Levi-Civita, which clearly is a very special choice. It is symmetric and metric-compatible, i.e. $\nabla_\alpha g_{\mu\nu}=0$. When the latter condition is violated, we have {\it non-metricity}, which in general corresponds to $D^2(D+1)/2$ independent degrees of freedom in a $D$-dimensional spacetime. If the connection has an antisymmetric component, we are said to have {\it torsion}, which adds $D^2(D-1)/2$ independent degrees of freedom. 

Recently it has been proposed that the spacetime connection could be generated by an independent metric \cite{Koivisto:2011vq}. Then, if we call this new metric $\hat{g}_{\mu\nu}$, it acts as the potential for the spacetime connection as  
\be \label{l-s-hat}
\hat{\Gamma}^{\al}_{\phantom{\al}\bt\ga}=\frac{1}{2}\hat{g}^{\al\la}\lp \hat{g}_{\bt\la,\ga}+ \hat{g}_{\la\ga,\bt}-\hat{g}_{\bt\ga,\la}\rp\,.
\ee
In the C-theories of gravitation \cite{Amendola:2010bk,Koivisto:2011tp} a conformal relation is assumed between the two metrics, $\hat{g}_{\mu\nu} = C(\R){g}_{\mu\nu}$, where $\R=g^{\mu\nu}\hat{R}_{\mu\nu}(\hat{\Gamma})$ is the curvature of \jose{the} spacetime. The C-theories introduce completely new theories and unify Einstein gravity 
and the so called Palatini gravity which correspond to specific choices of the conformal relation $\jose{C(\R)}$. To sum up, Einstein's GR imposes the
a priori rule (\ref{l-c}) for the connection, and the C-theories are a conformal generalization of this rule. An alternative approach is not to prescribe a rule for the connection, but consider the connection-generator $\hat{g}_{\mu\nu}$ as an independent tensor field. This approach we 
call the (unconstrained) {\it bimetric variational principle}. There the conformal degree of freedom adds effectively a massless scalar field \cite{Koivisto:2011vq}. This already shows that the resulting theory is physically distinct from metric GR, but what happens to the other 
degrees of freedom remains to be clarified. This is what we pursue in this study. 

However, it is first useful to briefly comment upon the difference of our new approach to the standard metric-affine variational principle \cite{Hehl:1994ue}. The latter,
often called the Palatini variation, promotes the connection into a fundamental field. It is well known that for the Einstein-Hilbert Lagrangians
coupled to nonspinning matter, the ''Palatini's device'' results in GR\jose{\footnote{\jose{To be precise, one needs to assume further conditions on the connection to uniquely recover GR in the Palatini formalism (see for instance \cite{Clifton:2011jh}).}}}, whereas for many classes of Lagrangians it generates underdetermined 
field equations \cite{0305-4470-12-8-017,Ferraris:1992dx}. More recent studies have discovered more problematical aspects of the resulting theories\footnote{Some of the problems are cured when one considers a hybrid metric-Palatini theory \cite{Harko:2011nh}.} \cite{Koivisto:2005yc,Iglesias:2007nv,Barausse:2007pn,Koivisto:2010jj}, though they have some attractive properties in view of cosmology and even quantum gravity phenomenology, 
see the review \cite{Olmo:2011uz}. From different viewpoints, the culprit for the problems has been identified as the nontensorial
nature of the connection field \cite{Amendola:2010bk}, the mixing of the metric-affine and Riemannian frameworks \cite{Koivisto:2005yk,Goenner:2010tr} and the lack of dynamics \cite{capozziello2010beyond}. Adopting the bimetric variational principle removes all these possible issues. 

In section \ref{bime} we will derive the connection from a general tensor potential up to the leading order. In section \ref{a-sector}
we shall analyze the antisymmetric sector of the theory. There appears an interesting model with propagating torsion. In section \ref{s-sector} we analyze the symmetric sector of the theory, and find that when applied to an action that is linear in the curvature, there appear additional
ghost degrees of freedom. This rules out the bimetric variational principle when applied to the simple Einstein-Hilbert -like action. 
This leads us to consider a deformation of this action. Indeed, in section \ref{symm} we show that, by adding a standard Einstein-Hilbert term, the transverse traceless ghosts can be exorcised, although \mk{then} the conformal mode becomes  problematic. We conclude in section \ref{conc}. 

\mk{Throughout the paper a "mostly plus" signature of the metric is used $(-,+,+,+)$.}

\section{The Bimetric variational principle}
\label{bime}

Our starting point is to imagine $\hat{g}_{\mu\nu}$ such that we need not have $\hGamma^\alpha_{[\mu\nu]}=0$, while according to (\ref{l-s-hat}) we have 
\be
\hat{\nabla}_\alpha \hat{g}_{\mu\nu}=0\,.
\label{bgcomp}
\ee 
The field $\hat{g}_{\mu\nu}$ should be nondegenerate in order to have a well-defined inverse \jose{$\hat{g}^{\mu\nu}$}. Since this field is nothing but the potential for the geometric connection, we do not need to impose further constraints on it. In particular, since it does not correspond to physical distances, it does not need to be symmetric $\hat{g}_{\mu\nu}\neq \hat{g}_{\nu\mu}$.
We are thus led to consider the action
 \be \label{action}
 S=\int d^4x \sqrt{-g}\left[\frac{1}{16\pi G}g^{\mu\nu}\hR_{\mu\nu}\lb\hGamma(\hat{g}_{\mu\nu})\rb+\Lag_m[\psi_\alpha,\hat{\nabla}\psi_\alpha, g_{\mu\nu}]\right],
 \ee
where $\psi_\alpha$ denotes the matter fields. Apart from a couple of hats added, this action is just the usual Einstein-Hilbert action. However, the differences are profound, as we shall find in the following. In order to proceed further we shall split the metric potential into its symmetric and antisymmetric components, namely:
\be
\hat{g}_{\mu\nu}=\bg_{\mu\nu}+\Sigma_{\mu\nu}
\ee
with $\bg_{\mu\nu}=\hat{g}_{(\mu\nu)}$ and $\Sigma_{\mu\nu}=\hat{g}_{[\mu\nu]}$ and we assume that they are not degenerate. Then, if we insert this decomposition into (\ref{bgcomp}) we obtain the two sets of equations:
\begin{eqnarray}
\hat{\nabla}_\alpha \bg_{\mu\nu}=0,\label{bgeq}\\
\hat{\nabla}_\alpha \Sigma_{\mu\nu}=0.\label{Seq}
\end{eqnarray}
These equations determine the connection $\hGamma^\alpha_{\beta\gamma}$ in terms of $\bg_{\mu\nu}$ and $\Sigma_{\mu\nu}$. Now, in order to give a more intuitive view of the theory we shall also split the connection as
\be \label{hgamma}
\hGamma^\alpha_{\beta\gamma}=\bGam^\alpha_{\beta\gamma}+S^{\alpha}_{\beta\gamma}
\ee
with $\bGam^\alpha_{\beta\gamma}=\hGamma^\alpha_{(\beta\gamma)}$ and $S^\alpha_{\beta\gamma}=\hGamma^\alpha_{[\beta\gamma]}$. Then, Eq. (\ref{bgeq}) and (\ref{Seq}) will give us $\bGam$ and $S$ in terms of $\bg$ and $\Sigma$. This completes the exposition of the bimetric variational principle. It
is clear we can decompose the fundamental degrees of freedom in gravity in this set-up to the usual metric \jose{plus} the symmetric and the antisymmetric
pieces of the connection-generating potential. We shall investigate each of the latter sectors separately.

In the following we shall restrict to perturbative order. A perturbative solution to the equations above is given by \cite{Valkenburg}:
\begin{eqnarray}
\bGam^{\alpha}_{\beta\gamma}&=&\frac{1}{2}\bg^{\lambda\alpha}\left(\partial_\gamma\bg_{\beta\lambda}+\partial_\beta\bg_{\lambda\gamma}-\partial_\lambda\bg_{\beta\gamma}\right),\\
S^{^\lambda}_{\beta\rho}&=&\frac{1}{2}\bg^{\lambda\alpha}\left(\bnabla_\beta\Sigma_{\alpha\rho}+\bnabla_\rho\Sigma_{\beta\alpha}+\bnabla_\alpha\Sigma_{\beta\rho}\right).\label{Ssigma}
\end{eqnarray}
where $\bnabla$ is the covariant derivative for the connection $\bGam$.  This perturbative solution has been obtained at first order in $\Sigma$, assuming that it is small as compared to the symmetric part. 

The bimetric variational principle proposed here differs from other bimetric theories of gravity in several aspects. Concerning the theories in which one of the fields play the role of background metric and it is non-dynamical, the difference is clear, since we have that the two metrics are dynamical. Thus, our variational principle could only be equivalent to bimetric theories in which the two fields are dynamical. Moreover, one of our fields carry a non-symmetric component, so it could only be equivalent to bimetric theories with torsion. In any case, as we shall see later, when the antisymmetric part of $\hat{g}_{\mu\nu}$ is small, it decouples from the symmetric component at quadratic level and one could expect the resulting action for the symmetric component to be equivalent to existing bimetric theories. To show that this is not necessary the case,  let us consider the usual theories with two dynamical metrics $f_{\mu\nu}$ and $g_{\mu\nu}$ that interact through a certain potential described by the action:
\be
S[f_{\mu\nu},g_{\mu\nu}]=\frac{1}{M_f^2}\int d^4x\sqrt{-f} R(f)+\frac{1}{M_g^2}\int d^4x\sqrt{-g} R(g)+S_{int}[f_{\mu\nu},g_{\mu\nu}].
\ee
In this theory, the interaction term contains non-derivative couplings between the two metrics. However, in our proposed bimetric variational principle, the coupling between the two spin-2 fields is only through derivative interactions. In particular, it seems apparent that one cannot recover the Fierz-Pauli action for a massive spin-2 field from our variational principle for the linearized theory as one would require for $S_{int}$. We shall explicitly show this below. Thus, we do not expect the results obtained in those theories to be directly applicable to our case. 

Another difference appearing in our bimetric variational principle is that the field $\hat{g}_{\mu\nu}$ only appears as an {\it auxiliary} field or, in other words, it is a tensor potential for the connection of the spacetime. In that sense, this field is not {\it physical} and only the curvature generated by it can have physical effects, i.e., it can only appear through its associated curvature. This is crucially different from other bimetric theories because there both metrics are physical. This can be used as a guiding principle to add more terms in the action since $\hat{g}$ cannot appear directly, but only through $\hat{R}^\alpha_{\;\;\beta\gamma\delta}$ and its contractions with the spacetime metric $g_{\mu\nu}$.

\section{The antisymmetric sector}
\label{a-sector} 

An interesting consequence of the restored dynamics is propagating torsion. 
At the classical level, torsion proves to be essential for total angular momentum conservation when intrinsic spin angular momentum is 
relevant (for reviews on torsion, 
see \cite{Shapiro:2001rz,Hammond:2002rm}). 
On the other hand, at the quantum level, a covariant version of Dirac equation also requires the presence of a torsion. This necessity 
roots on the {\it mass-reversal} symmetry
\footnote{The mass reversal symmetry of the Dirac equation for the spinor field $\psi$ with mass $m$ consists of the combined transformations: $\psi\rightarrow e^{i\alpha\gamma^5}\psi$ and $m\rightarrow-m$.} 
of the Dirac equation when we want 
\mk{it to be} a local symmetry, as required by special relativity. This could arguably be seen as one of the indications 
that torsion {\it must} be present in a fundamental theory of gravity, as discussed in \cite{springerlink:10.1007/s10714-010-1045-x}. As 
we will show below and contrary to the metric-affine variation\footnote{The metric approach can be rendered equivalent by a slight change 
of the Lagrangian \cite{PhysRev.77.699}.}, the bimetric variation procedure naturally yields a dynamical torsion field.

Let us now consider the decomposition in (\ref{hgamma}). As one might expect, the symmetric component of the connection is nothing but the Levi-Civita connection of $\bg_{\mu\nu}$, i.e.,\footnote{The reader should be careful not to get confused between $\hat{\nabla}$ and $\bnabla$.} $\bnabla_\alpha \bg_{\mu\nu}=0$  and the torsion field is determined by the antisymmetric component of the metric potential. An important constraint satisfied by $S^\alpha_{\beta\gamma}$ is that it is traceless $S^\alpha_{\beta\alpha}=0$, as can be directly seen from the above expression. 
Then, it is straightforward to write down the Ricci tensor in terms of $\bGam^\alpha_{\beta_\gamma}$ and $S^\alpha_{\beta_\gamma}$:
\begin{eqnarray}
\hR_{\mu\nu}(\hGamma)=\hR_{\mu\nu}(\bGam)+\partial_\alpha S^\alpha_{\mu\nu}+\bGam^\alpha_{\alpha\beta}S^\beta_{\mu\nu}-2\bGam^\alpha_{\beta[\nu}S^\beta_{\mu]\alpha}+S_{\mu\alpha}^\lambda S_{\nu\lambda}^\alpha
\end{eqnarray}
where we have used the traceless  property of the torsion tensor. Thus, when we plug this expression into the action, we obtain the following:
\be
S=\frac{1}{16\pi G}\int d^4x\sqrt{-g}g^{\mu\nu}\left[\hR_{\mu\nu}(\bGam)+S_{\mu\alpha}^\lambda S_{\nu\lambda}^\alpha\right].\label{effecact}
\ee
Once again, this action might seem not to be new in the sense that it reproduces the usual Einstein-Hilbert action in the presence of torsion. However, there is a crucial difference. 
The torsion field itself is not the fundamental field, but it must be expressed in terms of $\bg$ and $\Sigma$, since it is them where the
antisymmetric part of the spacetime connection emerges from. Thus, the fundamental fields of this theory are $g_{\mu\nu}$, $\bg_{\mu\nu}$ and $\Sigma_{\alpha\beta}$. At first sight, the true spacetime metric tensor $g_{\mu\nu}$ might seem not to be dynamical because there is no kinetic term for it. This, however, turns out to be wrong because the equations of motion for $\bg_{\mu\nu}$ will actually give evolution equations for $g_{\mu\nu}$. Another interesting feature of the above action is that, since $S^\alpha_{\mu\nu}$ depends on first derivatives of $\Sigma_{\alpha\beta}$, we indeed obtain a propagating torsion field even for the simple Einstein-Hilbert term. Indeed, Eq.  (\ref{Ssigma}) can be recast in the following form:
\be
S^\alpha_{\beta\gamma}=\bg^{\alpha\sigma}\left(\frac{1}{2}F_{\sigma\gamma\beta}-\bnabla_\sigma\Sigma_{\gamma\beta}\right),
\label{SFdef}
\ee
where $F_{\alpha\beta\gamma}$ is the strength field tensor of $\Sigma_{\alpha\beta}$ (Kalb-Ramond field), so that the gravitational action can also be written as:
\begin{eqnarray}
S=\int d^4x\sqrt{-g}\frac{g^{\mu\nu}}{16\pi G}\left[\hR_{\mu\nu}(\bGam)
+\bg^{\alpha\sigma}\bg^{\lambda\rho}\left(\frac{1}{2}F_{\sigma\lambda\mu}-\bnabla_\sigma\Sigma_{\lambda\mu}\right)\left(\frac{1}{2}F_{\rho\alpha\nu}-\bnabla_\rho\Sigma_{\alpha\nu}\right)\right],
\end{eqnarray}
where the dynamical origin of the torsion field becomes apparent. If we assume that the antisymmetric component of the connection potential $\Sigma_{\alpha\beta}$ can be derived from a potential, i.e., $\Sigma={\rm d} \sigma$, then $F=0$ and we obtain a gauge-fixing-like term for $\Sigma$. However, if we promote $\sigma$ into the fundamental field instead of $\Sigma$, we will obtain higher order equations of motion that could potentially introduce Ostrogradsky instabilities. \jose{Another possibility would be to make the torsion field be proportional to the Kalb-Ramond field so that the effective action becomes nothing but the standard Faraday action for a 2-form. }The stability of the antisymmetric \jose{part} may indeed require us to impose this constraint \cite{VanNieuwenhuizen:1973fi}. This could for instance be achieved by imposing a gauge symmetry on the torsion sector, as it happens for instance when one assumes that the torsion is fully derived from a tensor potential
  \cite{Hammond:1991ty}.  In any 
case, stability of each particular possibility should be carefully studied.

\section{The symmetric sector}
\label{s-sector}

Next we study the stability of the symmetric piece. For this purpose, in the following we shall consider the perturbed theory around GR so that:
\be
{\bar g}_{\mu\nu}=g_{\mu\nu}+{\bh}_{\mu\nu}
\ee
with $\bh_{\mu\nu}$ a small perturbation. Since we focus on the symmetric sector in this section, let us neglect the contribution from 
$\Sigma_{\mu\nu}$.
This allows us to extend some of the results beyond linear perturbations.

To proceed, we first notice that the inverse metric can be found as 
\be
\label{inverse}
\bar{g}^{\mu\nu}=\left[\left(g+\bh\right)^{-1}\right]^{\mu\nu}=
\left[\left({\mathbb I}+g^{-1}\bh\right)^{-1}\right]^{\mu}_{\alpha}g^{\alpha\nu}=
\left[\sum_{n=0}^{\infty}\left(-g^{-1}\bh\right)^n\right]^{\mu}_{\alpha}g^{\alpha\nu}
\ee
where $\left(-g^{-1}\bh\right)^0\equiv {\mathbb I}$, the unit matrix, and demanding
that the series be convergent gives a precise meaning to the smallness of $\bh$.

Now, the connection coefficients read
\be
\bar{\Gamma}^{\mu}_{\alpha\beta}=\frac12 \bar{g}^{\mu\nu}\left(\partial_{\alpha}\bar{g}_{\nu\beta}+
\partial_{\beta}\bar{g}_{\nu\alpha}-\partial_{\nu}\bar{g}_{\alpha\beta}\right)=
\frac12 \left[\sum_{n=0}^{\infty}\left(-g^{-1}\bh\right)^n\right]^{\mu}_{\rho}
\left(2\Gamma^{\rho}_{\alpha\beta}+g^{\rho\nu}\left(\partial_{\alpha}\bh_{\nu\beta}+
\partial_{\beta}\bh_{\nu\alpha}-\partial_{\nu}\bh_{\alpha\beta}\right)\right).
\ee
At the $0$-th order in $\bh$ it gives, of course, $\Gamma^{\mu}_{\alpha\beta}$. At the
$n$-th order, $n\geq1$, one has to use the n-th term of the sum with the first
term in the brackets on the right, and the $(n-1)$-th term in the sum with the
second term in the brackets. It yields
\begin{eqnarray}
\label{hatconnection}
\bar{\Gamma}^{\mu}_{\alpha\beta}&=&\Gamma^{\mu}_{\alpha\beta}+
\left[\sum_{n=0}^{\infty}\left(-g^{-1}\bh\right)^n\right]^{\mu}_{\rho}
\left(\frac12 g^{\rho\nu}\left(\partial_{\alpha}\bh_{\nu\beta}+
\partial_{\beta}\bh_{\nu\alpha}-\partial_{\nu}\bh_{\alpha\beta}\right)-
g^{\rho\nu}\bh_{\nu\gamma}\Gamma^{\gamma}_{\alpha\beta}\right)\\
&=&\Gamma^{\mu}_{\alpha\beta}+\frac12
\left[\sum_{n=0}^{\infty}\left(-g^{-1}\bh\right)^n\right]^{\mu}_{\rho}
g^{\rho\nu}\left(\nabla_{\alpha}\bh_{\nu\beta}+
\nabla_{\beta}\bh_{\nu\alpha}-\nabla_{\nu}\bh_{\alpha\beta}\right).
\end{eqnarray}
An interesting feature is that
\be
\delta\Gamma^{\mu}_{\alpha\beta}=\frac{1}{2}\bar{g}^{\mu\nu}\left(\nabla_{\alpha}\bh_{\nu\beta}+
\nabla_{\beta}\bh_{\nu\alpha}-\nabla_{\nu}\bh_{\alpha\beta}\right)
\ee
where the matrix $\bar{g}$ should be regarded as a function of matrices $g$ and $\bh$. From the last expression we see that
\be
\delta\Gamma^{\mu}_{\alpha\mu}=\frac{1}{2}\bar{g}^{\mu\nu}\nabla_{\alpha}\bh_{\mu\nu}
\ee
due to the symmetry of the metric $\bar{g}^{\mu\nu}$. It also holds true at the linear
level in $\bh$ because the zeroth order approximation to $\bar{g}$ is equal to $g$ which is also
symmetric. Note that, generically, if we approximate the series for $\hat{g}$ only by terms
with $0\leqslant n\leqslant N$, then the sum would not be symmetric. However, its asymmetry is always
next order in $\bh$. Indeed,
\begin{eqnarray}
\left[\sum_{n=0}^{N}\left(-g^{-1}\bh\right)^n\right]^{\mu}_{\alpha}g^{\alpha\nu}&=&
\left[\left[\sum_{n=0}^{\infty}\left(-g^{-1}\bh\right)^n\right]^{\mu}_{\alpha}-
\left[\sum_{n=N+1}^{\infty}\left(-g^{-1}\bh\right)^n\right]^{\mu}_{\alpha}\right]g^{\alpha\nu}\\
&=&\left[{\mathbb I}-\left(-g^{-1}\bh\right)^{N+1}\right]
\left[\sum_{n=0}^{\infty}\left(-g^{-1}\bh\right)^n\right]g^{-1}=
\left[{\mathbb I}-\left(-g^{-1}\bh\right)^{N+1}\right]^{\mu}_{\alpha}g^{\alpha\nu}.
\end{eqnarray}
Next, it is necessary to evaluate the Ricci tensor,
\be
\label{hatRicci}
\bar{R}_{\mu\nu}=\partial_{\alpha}\bar{\Gamma}^{\alpha}_{\mu\nu}-
\partial_{\nu}\bar{\Gamma}^{\alpha}_{\mu\alpha}+
\bar{\Gamma}^{\alpha}_{\beta\alpha}\bar{\Gamma}^{\beta}_{\mu\nu}-
\bar{\Gamma}^{\alpha}_{\mu\beta}\bar{\Gamma}^{\beta}_{\nu\alpha}.
\ee
Substituting $\bar{\Gamma}^{\alpha}_{\mu\nu}=\Gamma^{\alpha}_{\mu\nu}+\delta\Gamma^{\alpha}_{\mu\nu}$,
where the \mk{\emph{tensor}} quantity $\delta\Gamma^{\alpha}_{\mu\nu}$ is already known \mk{from eq.~}(\ref{hatconnection}) in the form of
convergent (for small enough $\bh$) power series, we easily get
\be
\label{varRicci}
\bar{R}_{\mu\nu}=R_{\mu\nu}+
\nabla_{\alpha}\delta\Gamma^{\alpha}_{\mu\nu}-
\nabla_{\nu}\delta\Gamma^{\alpha}_{\mu\alpha}+
\delta\Gamma^{\alpha}_{\beta\alpha}\delta\Gamma^{\beta}_{\mu\nu}-
\delta\Gamma^{\alpha}_{\mu\beta}\delta\Gamma^{\beta}_{\nu\alpha}.
\ee
The derivative part of the Ricci tensor variation is purely a surface term, and can be neglected.
\footnote{Nevertheless, those derivatives can also be presented in a nice convergent form. The only ingredient we need is the covariant derivative of the power series
$$\sum_{n=0}^{\infty}\left(-g^{-1}\bh\right)^n\equiv\sum_{n=0}^{\infty}\left(-\mathfrak{h}\right)^n$$
where the $\mathfrak{h}$-matrix simply denotes the $\bh^{\mu}_{\nu}$-field. It is easy to see that differentiating the $\left(-\mathfrak{h}\right)^n$ term one gets $n$ terms with $-\nabla \mathfrak{h}$ in all possible positions. If we differentiate the whole sum, but consider only the terms with $(-\mathfrak{h})^m$ to the right from $-\nabla \mathfrak{h}$, then there would be exactly one possible term with any fixed power of $-\mathfrak{h}$ to the left, where $\left(-\mathfrak{h}\right)^k$ comes from the differentiation of $\left(-\mathfrak{h}\right)^{m+k+1}$
term in the initial sum. Therefore, it is easy to see that
$$\nabla_{\mu}\left[\sum_{n=0}^{\infty}\left(-\mathfrak{h}\right)^n\right]^{\alpha}_{\beta}=
-\left[\sum_{m=0}^{\infty}\left(-\mathfrak{h}\right)^m\right]^{\alpha}_{\rho}
\left(\nabla_{\mu} \bar{h}^{\rho}_{\sigma}\right)
\left[\sum_{k=0}^{\infty}\left(-\mathfrak{h}\right)^k\right]^{\sigma}_{\beta}.$$
Obviously, this formula generalises the differentiation rule
for a function $f(x)=\left(1+g(x)\right)^{-1}$.} 
Then, the resulting action is:
\begin{eqnarray}
\label{exactaction}
S&=&\jose{\frac{1}{16\pi G}}\int d^4 x \sqrt{-g}\left(R+g^{\mu\nu}\left(
\delta\Gamma^{\alpha}_{\beta\alpha}\delta\Gamma^{\beta}_{\mu\nu}-
\delta\Gamma^{\alpha}_{\mu\beta}\delta\Gamma^{\beta}_{\nu\alpha}\right)\right)=\jose{\frac{1}{16\pi G}}\int d^4 x \sqrt{-g}R \nonumber \\
&&+ \jose{\frac{1}{16\pi G}} \int d^4 x \frac{\sqrt{-g}}{4}
\left[\sum_{n=0}^{\infty}\left(-g^{-1}\bh\right)^n\right]^{\alpha}_{\kappa}
g^{\kappa\rho}\left[\sum_{m=0}^{\infty}\left(-g^{-1}\bh\right)^m\right]^{\beta}_{\gamma}
g^{\gamma\sigma}\times \nonumber \\ 
&&\left[\nabla_{\beta}\bh_{\alpha\rho}
\left(2\nabla_{\mu}\bh_{\sigma}^{\mu}-\nabla_{\sigma}\bh_{\mu}^{\mu}\right)
-\left(\nabla_{\mu}\bh_{\beta\rho}+
\nabla_{\beta}\bh_{\mu\rho}-\nabla_{\rho}\bh_{\mu\beta}\right)g^{\mu\nu}
\left(\nabla_{\nu}\bh_{\alpha\sigma}+
\nabla_{\alpha}\bh_{\nu\sigma}-\nabla_{\sigma}\bh_{\alpha\nu}\right)\right].
\end{eqnarray}
We see that the added ingredient is just a symmetric tensor field $\bh$ 
possessing a quadratic in $\nabla \bh$ action,
although with non-linear in $\bh$ non-derivative coefficients.
At the first (quadratic)
order the coefficient in front of the covariant derivatives reduces simply
to $\frac14 g^{\alpha\rho}g^{\beta\sigma}$. 
(In general, this coefficient is equal to
$\frac14 \bar{g}^{\alpha\rho}\bar{g}^{\beta\sigma}$ as a function
of $g$ and $\bh$. And this is actually a non-perturbative result as long as $\Sigma_{\mu\nu}$ vanishes, valid at
any, not necessarily small, values of $\bh$.)
In this limit the action \mk{in eq.~}(\ref{exactaction}) is just equivalent to the linearized Einstein-Hilbert action:
\be
\label{linaction}
S=\jose{\frac{1}{16\pi G}}\int d^4 x \sqrt{-g}\left[R+\frac14\left(\partial_{\alpha}\bh_{\mu\nu}\partial^{\alpha}\bh^{\mu\nu}-
2\partial^{\alpha}\bh_{\mu\nu}\partial^{\nu}\bh^{\mu}_{\alpha}+
2\partial_{\alpha}\bh^{\alpha\mu}\partial_{\mu}\bh^{\beta}_{\beta}-
\partial_{\mu}\bh^{\alpha}_{\alpha}\partial^{\mu}\bh^{\beta}_{\beta}\right)\right]
\ee
where we have assumed that the metric $g_{\mu\nu}$ is the Minkowski one,
and therefore all covariant derivatives
are substituted by the ordinary ones.

One of the two gravitons has inevitably ghost excitations. \jose{Indeed, we were assuming the ${\rm diag} (-,+,+,+)$-signature of the metric throughout the paper which implies that the $\bh$-field is a tensor ghost. Of course, if we switch to the opposite sign convention, then the $\bh$-field is healthy. But after separating it, we have the GR action $\propto\int d^4 x \sqrt{-g}R$ while, with the new convention, it must have an opposite sign.} Let us \mk{demonstrate} the problem explicitly \mk{by expanding the metric $g_{\mu\nu}$} around the Minkowski space\mk{time}
$g_{\mu\nu}=\eta_{\mu\nu}+ h_{\mu\nu}$. The second order action for $h$ acquires
the form
\be
\label{actvar}
S^{(2)}=\jose{\frac{1}{16\pi G}}\int d^4 x\left[ \sqrt{-g}g^{\mu\nu}(\delta^{(2)}R_{\mu\nu})+ 
(\delta^{(1)}(\sqrt{-g}g^{\mu\nu}))(\delta^{(1)}R_{\mu\nu})\right].
\ee
The first term we already know \mk{from eq.~}(\ref{varRicci}), $\delta^{(2)}R_{\mu\nu}=
\delta\Gamma^{\alpha}_{\beta\alpha}\delta\Gamma^{\beta}_{\mu\nu}-
\delta\Gamma^{\alpha}_{\mu\beta}\delta\Gamma^{\beta}_{\nu\alpha}+{\rm surface\ terms}$,
and it gives precisely the same action \mk{for $h$ as the action for $\bh$ in eq.~\eqref{linaction}.}
However, it is not yet the end of the story. With $\delta^{(1)}R_{\mu\nu}=
\partial_{\alpha}\delta\Gamma^{\alpha}_{\mu\nu}-
\partial_{\nu}\delta\Gamma^{\alpha}_{\mu\alpha}$ 
and
$\delta^{(1)}(\sqrt{-g}g^{\mu\nu})=-h^{\mu\nu}+
\frac12 h^{\beta}_{\beta}\eta^{\mu\nu}$,
we write the second term in (\ref{actvar}) in the form
\be
\frac12 \left(-h^{\mu\nu}+\frac12 h^{\beta}_{\beta}\eta^{\mu\nu}\right)\left(
\partial^2_{\nu\alpha}h^{\alpha}_{\mu}+\partial^2_{\mu\alpha}h^{\alpha}_{\nu}-
{\partial^2}^{\alpha}_{\alpha}h_{\mu\nu}-\partial^2_{\mu\nu}h^{\alpha}_{\alpha}\right)
\ee
which, after some integration by parts, transforms to
\be
-\frac12 \left(\partial_{\alpha}h_{\mu\nu}\partial^{\alpha}h^{\mu\nu}-
2\partial^{\alpha}h_{\mu\nu}\partial^{\nu}h^{\mu}_{\alpha}+
2\partial_{\alpha}h^{\alpha\mu}\partial_{\mu}h^{\beta}_{\beta}-
\partial_{\mu}h^{\alpha}_{\alpha}\partial^{\mu}h^{\beta}_{\beta}\right)
\ee
and, together with the first part, it gives the final result:
\be
S^{(2)}=- \jose{\frac{1}{64\pi G}} \int d^4 x
\left[\partial_{\alpha}h_{\mu\nu}\partial^{\alpha}h^{\mu\nu}-
2\partial^{\alpha}h_{\mu\nu}\partial^{\nu}h^{\mu}_{\alpha}+
2\partial_{\alpha}h^{\alpha\mu}\partial_{\mu}h^{\beta}_{\beta}-
\partial_{\mu}h^{\alpha}_{\alpha}\partial^{\mu}h^{\beta}_{\beta}\right].
\ee
We see that the tensor modes $h^{(TT)}_{ik}$ are ghosts,  \jose{where $TT$ stands for spatially transverse and traceless.}

The total second order action for both types of perturbations acquires the form
\begin{align}
\label{ghosts}
S^{(2)}=\jose{\frac{1}{16\pi G}}\int d^4 x \left[\frac14\left(\partial_{\alpha}\bh_{\mu\nu}\partial^{\alpha}\bh^{\mu\nu}-
2\partial^{\alpha}\bh_{\mu\nu}\partial^{\nu}\bh^{\mu}_{\alpha}+
2\partial_{\alpha}\bh^{\alpha\mu}\partial_{\mu}\bh^{\beta}_{\beta}-
\partial_{\mu}\bh^{\alpha}_{\alpha}\partial^{\mu}\bh^{\beta}_{\beta}\right)\right.\nonumber \\
-\left.\frac14 \left(\partial_{\alpha}h_{\mu\nu}\partial^{\alpha}h^{\mu\nu}-
2\partial^{\alpha}h_{\mu\nu}\partial^{\nu}h^{\mu}_{\alpha}+
2\partial_{\alpha}h^{\alpha\mu}\partial_{\mu}h^{\beta}_{\beta}-
\partial_{\mu}h^{\alpha}_{\alpha}\partial^{\mu}h^{\beta}_{\beta}\right)\right].
\end{align}
At this level we have two separate diffeomorphism invariances (one of which is fake), and therefore can not judge upon the scalar
and vector degrees of freedom. However, we definitely see that the two types of tensor modes do
have opposite signs of kinetic functions. 

In summary, the theory is full of ghosts, at least in the simplest version of this class of models which
we have just considered. After all, it is not probably that much surprising. Let's forget about the tensor
nature of the metric fields. We take two scalar variables instead, $x$ for $\hat g$ and $y$ for $g$.
Suppose we have an action of the form $S=\int dt ({\ddot x} + {\dot x}^2)y$, somewhat analogous to our
gravity model. Integrating by parts we have $S=\int dt (-{\dot x}{\dot y}+{\dot x}^2 y)$. After
diagonalising this action we get $S=\int dt\ y \left(({\dot x}-\frac12 \frac{{\dot y}}{y})^2-
\jose{\frac14  \frac{{\dot y}^2}{y^2}}\right)=\int dt\ e^z\left[\left({\dot x}-\frac12 {\dot z}\right)^2
-\frac14 {\dot z}^2\right]$ where $z\equiv\log y$, and see that one of the modes is always a ghost.  Finally, it is interesting to note that no mass terms appear at this level for none of the metrics, as it was expected from our discussion above pointing out the differences of this theory with usual bimetric theories. The reason for the absence of the mass term is that, even though the fields will interact at higher orders, all the interactions will be through derivative couplings.

\section{On more elaborate theories}
\label{symm}

To render the theory viable, we have to reconsider our action.
A very simple way to extend the previous version of the theory is to consider the curvature $\R\jose{=g^{\mu\nu}\hat{R}_{\mu\nu}}$ to be added to the usual
Einstein-Hilbert term $\jose{R=g^{\mu\nu}R_{\mu\nu}}$. This way it loses some of the elegance, but we can use it as a proof of concept.
Let us say the relative coupling strength
of the nonmetric degrees of freedom is given by the parameter $0<\hat{\kappa}<1$. This adds one parameter to the theory
that can then in principle be constrained by the experimental tests of gravity. So, we write the action as
\be \label{ehb}
\jose{S =  \frac{1}{16\pi G} \int d^4 x \sqrt{-g}g^{\mu\nu}\lp  R_{\mu\nu}- \hat{\kappa}\hat{R}_{\mu\nu}\rp}\,. 
\ee
Using the calculations of the previous section it is easy to see that, at quadratic order and with Newton's constant renormalised by $G\rightarrow G/(1-\hat{\kappa})$, the action becomes:
\begin{eqnarray} \label{ehb2}
S^{(2)}&=&-\frac{1}{64\pi G}\int d^4 x\Big[  
\left(\partial_{\alpha}h_{\mu\nu}\partial^{\alpha}h^{\mu\nu}-
2\partial^{\alpha}h_{\mu\nu}\partial^{\nu}h^{\mu}_{\alpha}+
2\partial_{\alpha}h^{\alpha\mu}\partial_{\mu}h^{\beta}_{\beta}-
\partial_{\mu}h^{\alpha}_{\alpha}\partial^{\mu}h^{\beta}_{\beta}\rp \nonumber\\
&&+ \frac{\hat{\kappa}}{1-\hat{\kappa}} \lp\partial_{\alpha}\bh_{\mu\nu}\partial^{\alpha}\bh^{\mu\nu}-
2\partial^{\alpha}\bh_{\mu\nu}\partial^{\nu}\bh^{\mu}_{\alpha}+
2\partial_{\alpha}\bh^{\alpha\mu}\partial_{\mu}\bh^{\beta}_{\beta}-
\partial_{\mu}\bh^{\alpha}_{\alpha}\partial^{\mu}\bh^{\beta}_{\beta}\rp-\frac{4\hat{\kappa}}{1-\hat{\kappa}}S_{\mu\alpha}^\lambda S_{\nu\lambda}^\alpha
\Big]\,,
\label{kappaaction}
\end{eqnarray}
\jose{where now the two graviton fields are healthy as long as\footnote{\jose{This is true for the signature convention with ${\rm diag} (-,+,+,+)$.}} $0<\hat{\kappa}<1$}. That is, we obtain two copies of a massless spin-two fields plus the 2-form field determining the torsion of the geometric connection. It is remarkable that perturbations in both \jose{symmetric fields $h_{\mu\nu}$ and $\bh_{\mu\nu}$} happen to acquire the usual kinetic term of linearised GR, which is well-known to be precisely
the only viable action for pure massless spin-two fields\footnote{Though recently the result was generalized to nonlocal theories \cite{Biswas:2011ar}.} \cite{VanNieuwenhuizen:1973fi}. Thus, at the leading order, we have two sets of spin-2 fields entering into the action in a symmetric way. When matter is present, this symmetry is broken. Standard bosonic matter fields couple only to the spacetime geometry given by $g_{\mu\nu}$. This, for instance, is the case for scalar fields with standard kinetic terms or gauge (both abelian and non-abelian) fields constructed out of their strength fields\footnote{This statement depends on how we decide to covariantize the definition of the strength field. Let us consider the electromagnetic field defined in flat spacetime as $F_{\mu\nu}=\partial_\mu A_\nu-\partial_\nu A_\mu$. If we just replace partial derivatives by covariant derivatives, $F_{\mu\nu}$ will acquire a term involving torsion. However, if we define the strength field as the 
 exterior derivative of $A_\mu$ we 
do not have couplings to the connection.}. Of course, when non-minimal couplings are present, matter fields will also couple to $\hat{g}_{\mu\nu}$. This will be, for instance, the case of scalar fields \cite{Enqvist:2011qm} with higher order derivative terms, like K-essence or galileon fields, or vector fields with direct couplings to curvature. On the other hand, fermionic matter
like spinor fields are nonminimally coupled to the spacetime connection, and thus will always act as a source also for $\hat{g}_{\mu\nu}$. Since a Dirac field only couples to the totally antisymmetric part of the torsion field, such a coupling will be mediated by $F_{\alpha\beta\gamma}$, which is defined after \eqref{SFdef}. Of course, fermionic fields with higher spin will couple to the other components of the torsion field. Notice that, whereas nonminimal couplings of bosonic fields will lead to classical effecs, fermionic matter can only give rise to effects at the quantum level.
However, since the geometric connection has a non-symmetric part, i.e., torsion, one could couple this connection to fermionic fields, whereas bosonic fields would only couple to the matter connection. 
Moreover, the considered distinction between the metric and affine structures of the physical spacetime resolves the covariantization of the Dirac equation that leads to the need of a connection with torsion, as explained above. One may even speculate
on making contact with supersymmetry by a possible relation of $\hat{g}_{\mu\nu}$ with the supersymmetric partner of the graviton that couples
to fermions \cite{Deser:1976eh,Brink:1976sz}. 
   
To end, let us mention one important issue. The problematic part of quantizing Einstein gravity, and a big obstacle on the way of deforming it classically, is the conformal degree of freedom. There is indeed a scalar mode with the negative sign of kinetic
function. Fortunately for classical GR, it is killed by the gauge invariance. However, once you break it, you should expect some problems to come
about, such as the sixth mode or the Boulware-Deser ghost. In this bimetric context, we encounter the mirror image of this problem. Since the conformal degree of freedom for the action (\ref{action}) is a healthy scalar field \cite{Koivisto:2011vq}, it seems that by curing the other nonmetric degrees of freedom in the theory (\ref{ehb}), essentially by the minus sign in front of the coupling $\hat{\kappa}$, we have flipped the conformal mode into a ghost-like scalar. A simple way out would be a prescription where the conformal difference between the two metrics
is not allowed to propagate\footnote{Thus presenting a complete negative of the C-theories, where precisely the conformal relation is nontrivial.}.
Whether this is actually necessary in order to construct theories that are completely ghost-free to any order in perturbation theory, is a problem we cannot address within the scope of the present study. 

\section{Conclusions}
\label{conc}

The results presented here invite to study the nonlinear theory ensuing of bimetric variation. The main difficulty to go beyond the linear regime is to fully solve the equations (\ref{bgcomp}) that give the connection in terms of the potential metric fields expressed by means of $\bar{g}_{\mu\nu}$ and $\Sigma_{\mu\nu}$. Let us note that recently, in the context of
ghost-free massive gravity, the nonlinear couplings of two metrics that avoid the Boulware-Deser instability, have been discovered \cite{deRham:2010ik,deRham:2010k,Hassan:2011zd,Hassan:2011hr,Golovnev:2011aa}. Whether 
we can obtain those couplings from the bimetric variational principle remains to be explored. Also, because of the natural presence of propagating torsion in the considered bimetric formalism, some relations with {\it torsion massive gravity}, in which the torsion field provides the graviton with a mass \cite{Nair:2008yh,Deffayet:2011uk}, would be interesting to study further.

To conclude, in this paper we intended to present a new variational principle in which the affine connection is derived from a general tensor potential, which is different from the spacetime metric. We have studied the linear regime and found that the theory is equivalent to GR plus an additional pure spin-2 field and a 2-form leading to propagating torsion. However, the healthiness of the additional spin-2 field forced us to consider a more complicate action (\ref{ehb}) than the one consisting merely of a term linear in the curvature $\R$. Unfortunately, having cured the tensor modes, we rendered the conformal degree of freedom problematic. It remains to be seen whether it is possible to construct a completely viable theory in this formalism. In this respect, it is worth mentioning that the proposed bimetric theory is, in principle, different from those already existing in the literature so that the results about their stability cannot be directly applied to the present case.

\acknowledgments

TK thanks Stanley Deser and Nicola Tamanini for useful comments. 
AG and TK are grateful to the University of Granada for hospitality and partial
financial support during their visits.
TK is supported by the Norwegian Research council. JBJ is supported by the Ministerio de Educaci\'on under the postdoctoral contract EX2009-0305 and the spanish MICINN�s Consolider-Ingenio 2010 Programme under grant MultiDark CSD2009-00064. MK is supported by the grants CPAN CSD2007-00042 and MICINN (FIS2010-17395).

\bibliography{bimetric}

\end{document}